\documentclass[%
 reprint,
 aps,
 amsmath,amssymb,showpacs,showkeys,twocolumn
 r
]{revtex4-1}

\usepackage{graphicx}
\usepackage{dcolumn}
\usepackage{bm}
\usepackage{pstricks}
\usepackage{epstopdf}

\begin{document}

\preprint{APS/123-QED}

\title[]{Experimental demonstration of phase bistability in a broad area
optical oscillator with injected signal}

\author{R. Mart\'{\i}nez-Lorente,$^{1,*}$ A. Esteban-Mart\'{\i}n,$^2$ E. Rold\'{a}n,$^{1}$ K. Staliunas,$^3$ G. J. de Valc\'{a}rcel,$^{1}$ and F. Silva$^{1}$}

\address{$^1$Departament d'\`{O}ptica, Facultat de F\'{\i}sica, Universitat de Val\`{e}ncia, Dr. Moliner 50, 46100 Burjassot, Spain\\
$^2$ICFO---Institut de Ci\`{e}ncies Fot\`{o}niques, Mediterranean Technology
Park, 08860 Castelldefels, Spain \\
$^3$ICREA and Universitat Polit\`{e}cnica de Catalunya, Dept. de F\'{\i}sica i Enginyeria Nuclear, Colom 11, 08222 Terrassa, Spain \\
$^*$Corresponding author: ruben.martinez-lorente@uv.es}


\begin{abstract}
We demonstrate experimentally that a broad area laser-like optical
oscillator (a nondegenerate photorefractive oscillator) with structured
injected signal displays two-phase patterns. The technique (G. J. de Valc\'{a}rcel 
and K. Staliunas, Phys. Rev. Lett. \textbf{105}, 054101 (2010)) consists in
spatially modulating the injection, so that its phase alternates
periodically between two opposite values, i.e. differing by $\pi $. 
\end{abstract}


               
\maketitle

Bistability is a crucial mechanism for optical information encoding and
processing. When speaking of bistability one usually thinks of \textit{%
intensity} bistability, i.e. the stable coexistence of two states of unequal
field intensity, like the high and low transmission states of (so-called)
optical bistability \cite{OB1,OB2}, or the on and off states of optical
fiber solitons \cite{OFS1,OFS2} or of cavity solitons \cite%
{SCS1,SCS2,SCS3,SCS4,TCS1,TCS2,TCS3}. There is however an alternative type
of bistability, namely \textit{phase} bistability, in which two coexisting
stable states of equal intensity but opposite phase are supported by the
system. Phase bistability occurs in special nonlinear optical cavities, like
synchronoulsy pumped optical parametric oscillators \cite{SPOPO}. Phase-bistable states
are usually more symmetric comparing with amplitude-bistable states, it would
be desireable for optical information processing that lasers could display
such phase bistability.

The first proposal in that direction was given in \cite{Rock-original}, and
demonstrated in a laser-like system (specifically, a photorefractive
oscillator under nondegenerate wave mixing) in \cite{Rocktemporal-exp}: if
in a laser with injected signal the phase of driving field alternates
periodically in time between two opposite values (differing by $\pi $) at a
sufficiently high repetition rate, the phase of the slave laser can lock to
one of two possible values, both states having the same intensity. Such
driving technhique was termed "rocking" because, in a mechanical analogy,
that kind of injection tilts periodically (i.e. rocks) the laser potential
between two extreme positions \cite{Rock-original,Rock-classB,Rock-review}.
 Note that the phase of the emitted signal does not lock necessarily to either of the injection, 
say 0 and $\pi$, but rather to some values in between, which depends on the detuning between 
the injection and the cavity \cite{Rock-original, Rock-review}. The laser emission, 
simply speaking, avoids the action of the alternating injection, i.e. avoids the locations 
in phase space that are maximally affected by the alternating injection, and consequently moves to the most 
quiet locations. From the dynamical viewpoint such rocking is simillar 
to the stabilization of the topside position of the pendulum when the hanging point is vibrated 
in vertical direction (Kapitza pendulum).

This kind of rocking is however problematic in solid-state and
semiconductor lasers, the so-called class-B lasers, because the relaxation 
oscillations characterizing those lasers limit the performance of rocking \cite{Rock-classB}.
Consequently the initial concept of rocking in time was extended to the rocking in space
by considering the injection alternating in transverse space \cite{Spatialrock-theory, Spatialrock-exp}. 
The first proposal \cite{Spatialrock-theory} consisted of injecting a TEM$_{10}$ mode 
(displaying two opposite phases at the two mode lobes) into a low Fresnel number 
laser, capable to emit only on one, the lowest order, transverse mode. According to 
\cite{Spatialrock-theory} the phase of the slave laser locks to one among
two possible (and opposite) values, which was succesfully demonstrated in a
laser like oscillator \cite{Spatialrock-exp}. The concept was further
extended to broad area lasers \cite{Spatial-rocking}, in which many
transverse modes (a continuum of modes) play a role; it was predicted that 
under injection of a (monochromatic) beam displaying a spatial alternation of its phase between
two opposite values across its cross section, the emission of the slave
laser displays phase bistability. In this case, due to the spatially
extended nature of the system, different parts of the slave laser beam cross
section can take different phase values (among those special two), and phase
patterns are predicted to appear, opening the way in particular to phase
bistable cavity solitons \cite{Spatial-rocking}. Similarly to the rocking in time, where injection must vary sufficiently rapidly as compared to the characteristic time scale of the system (the cavity life-time), here in spatial rocking the injection must vary on a sufficiently small space scale as compared with the characteristic spatial scale of the system (diffraction length)\cite{Spatial-rocking}. Here we demonstrate experimentally the feasibility of this mechanism by using a photorefractive oscillator (PRO).

PROs are optical cavities containing a photorefractive crystal, which is
pumped by laser beams that do not resonate inside the cavity (e.g. because
they are tilted with respect to the cavity axis) \cite{Yeh,Yariv,Saleh}.
Under appropriate conditions (mainly crystal orientation and pump alignment)
an intracavity light field starts oscillating via efficient wave-mixing.
PROs are highly versatile systems for the study of nonlinear dynamics as
different wave mixings (two vs. four waves, or degenerate vs. nondegenerate)
can be tuned by using different resonators (ring or linear) and pumping
geometries (one pump or two counterpropagating pumps). In particular when
the cavity is linear and a single pump is used, the oscillation occurs due to a 
nondegenerate four wave mixing (NDFWM) process, and the phase of the self-generated
light is free, as in a free-running laser. However the similarities between lasers and
NDFWM PROs go far beyond that phase invariance: the PROs are laser-like
systems also from a nonlinear dynamics vewpoint \cite{PRO-laser}. In particular NDFWM 
PROs have been proven successful for the study of universal out of equilibrium pattern 
formation scenarios of phase invariant systems, such as vortex arrays and different traveling wave
patterns \cite{PRO-vortices}. 
\begin{figure}[tbh]
\centerline{\includegraphics[width=83 mm]{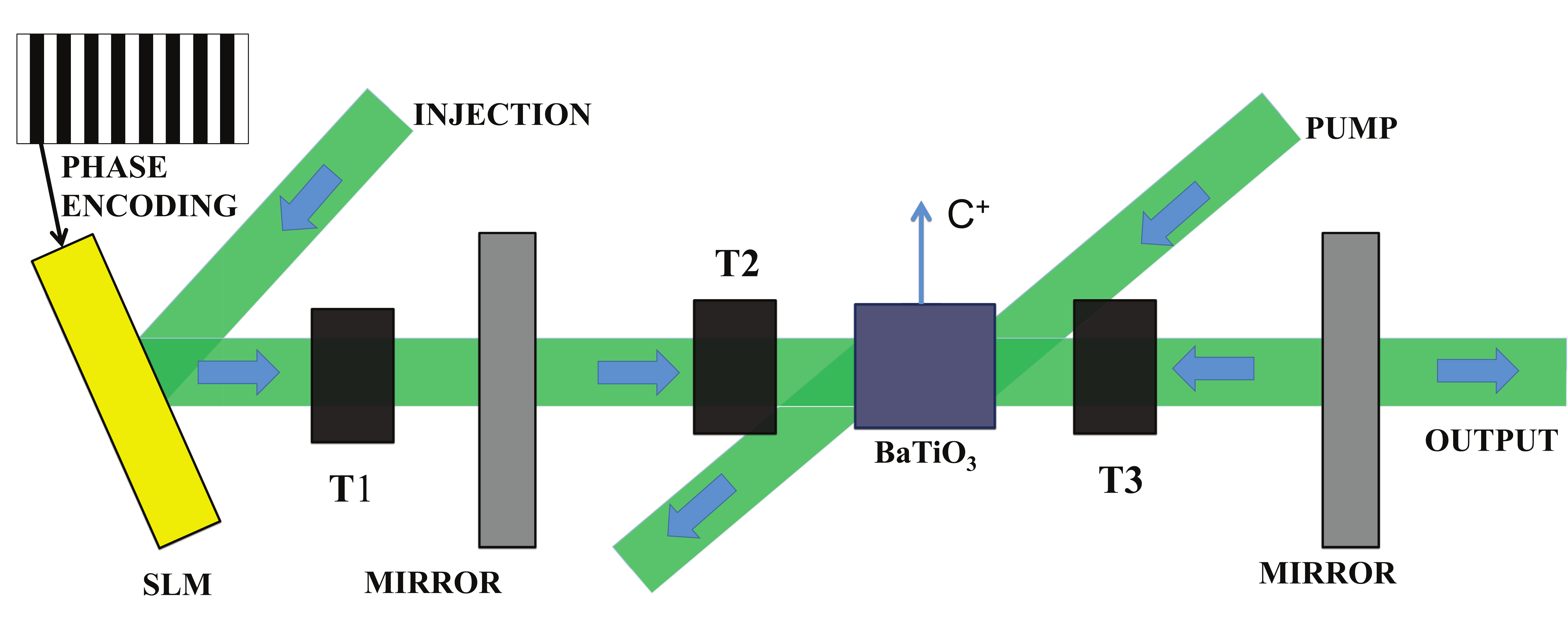}}
\caption{(Color online) Sketch of the photorefractive oscillator with injected signal
(rocking). Two mirrors form the resonator, which contains a BaTiO$_{3}$
crystal of 4.5$\times $4.5$\times $8.0 mm$^{3}$. SLM: spatial light
modulator. T1, T2, T3: telescopic systems.}
\label{scheme}
\end{figure}
In order to allow for pattern formation in the experiment the cavity Fresnel
number is made very large by two telescopic systems that image the cavity
mirrors close to the photorefractive BaTiO$_{3}$ crystal (nearly
self-imaging resonator \cite{Arnaud}); see Fig. \ref{scheme}. We inject a
"rocking beam" along the cavity axis, whose phase is tailored by means of a
spatial light modulator (SLM). The SLM (PLUTO-VIS-006-A, Holoeye Photonics
AG) is electrically addressed and is controlled by a computer in order to
give sharp $\pi $ phase jumps across the beam cross section. The rocking
beam, the pumping beam, and other auxiliary beams (for cavity length active
stabilization and for interferometry), all come from the same frequency
doubled Nd:YAG laser at 532 nm (Verdi V5, Coherent Inc.). For details about
the setup, the cavity stabilization procedure, for the interferometry, and
for the data processing, see \cite{APB}.

Owed to the small SLM pixel pitch ($8 \mu$m) large diffraction are observed on the reflected beam in such a way
that only the first two spatial harmonics enter efficiently into the cavity. The injection has then a sinusoidal variation across the resonator
transverse section, alternating periodically its sign (phase). 
\begin{figure}[tbh]
\includegraphics[width= 83 mm]{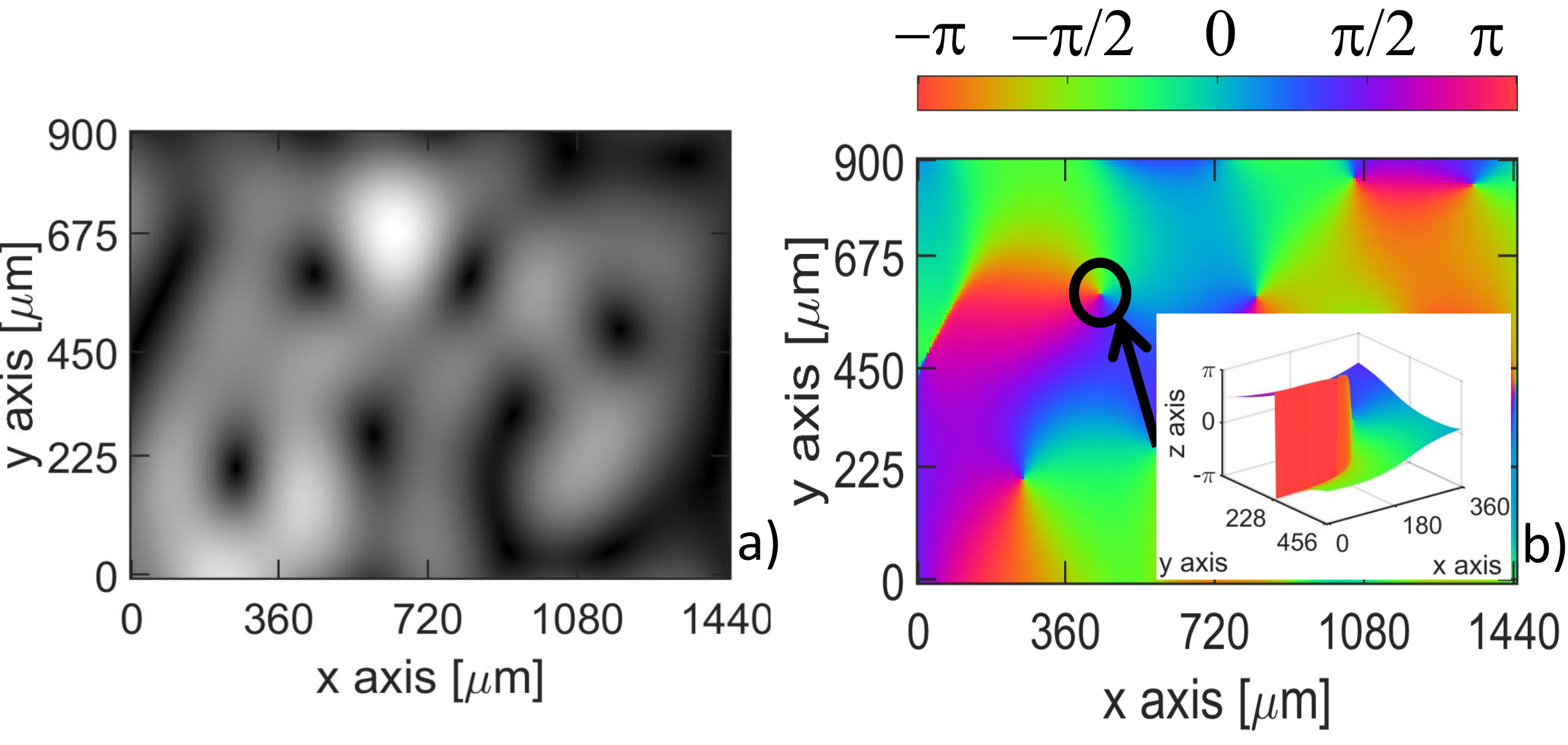} 
\caption{(Color online) Optical vortex array forming spontaneously in the free-running PRO.
Amplitude a) and phase b) maps of the field at the
nonlinear crystal. Spatial dimensions: 1440 $\mu$m (horizontal)$\times$ 900 $\mu$m (vertical).
A 3D representation of the phase around one of these optical vortices is shown in the inset in b),
with spatial dimensions 360 $\protect\mu$m$\times$450 $\mu $m; a
clear 2$\protect\pi $ rotation of the phase is appreciated.}
\label{vortices}
\end{figure}
This means that when the SLM is operated so that the phase changes by $\pi$ every 
three pixel rows (as is the case of the figures we show next), the effective rocking beam 
profile at the SLM has the 
form $\cos \left( 2\pi x/\Lambda_{\mathrm{SLM}}\right) $, with $\Lambda_{\mathrm{SLM}%
}=6\times 8\mu $m the spatial period. A telescope T1 images the SLM
plane onto the entrance cavity mirror, which in turn is imaged close to the
crystal by the left intracavity telescope T2 with total lateral
magnification of 0.5$\times $; hence the effective rocking beam profile at
the crystal has the form $\cos \left( 2\pi x/\Lambda_{\mathrm{C}}\right) $, with the
transverse period $\Lambda_{\mathrm{C}}=\cos(15^{\circ}) \times$ 24 $\mu$m = 23.2 $\mu$m 
approximately. Here we took into account the tilt of the SLM with respect the 
cavity axis.

\begin{figure}[tbh]
\includegraphics[width= 83 mm]{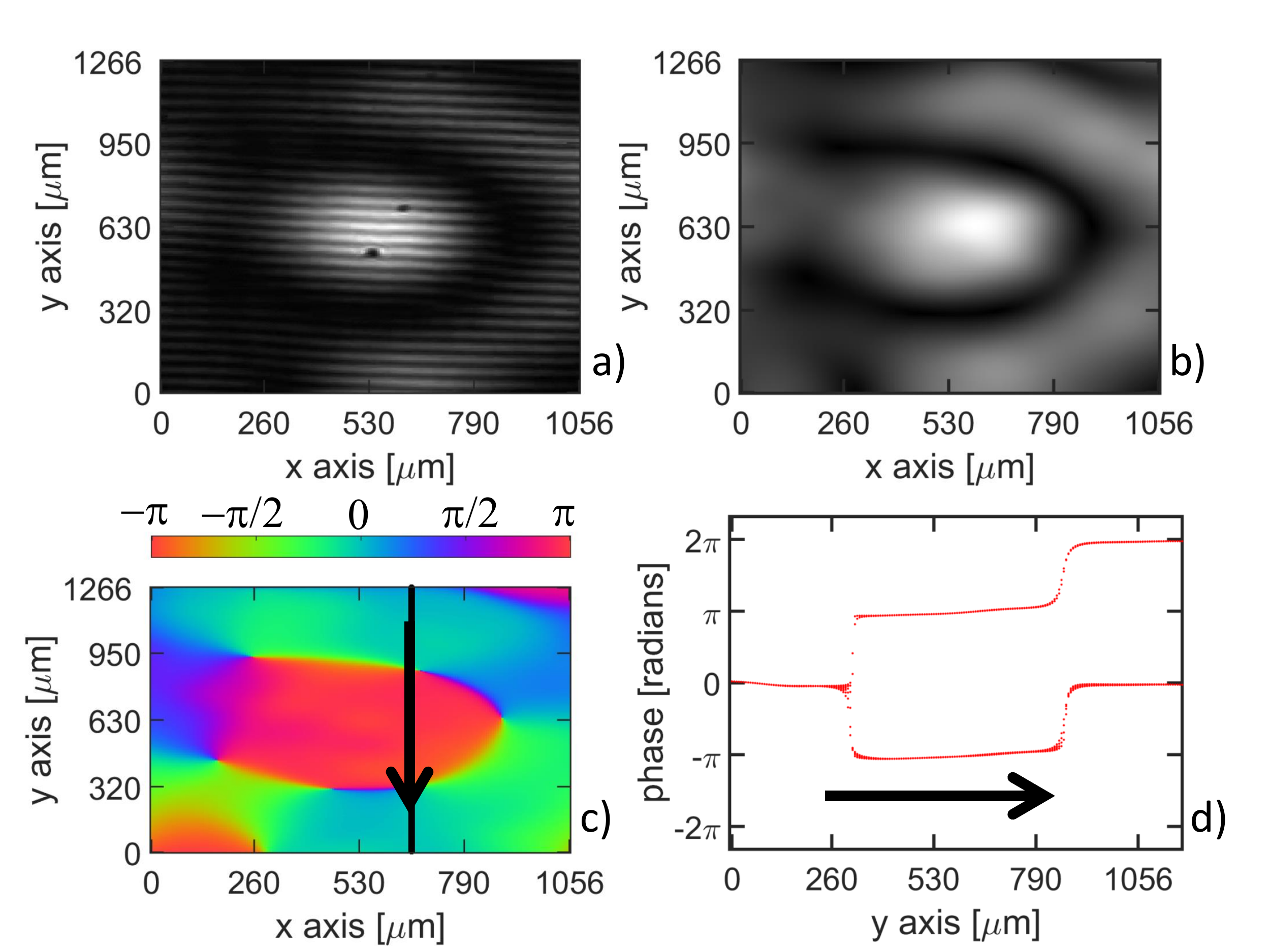}
\caption{(Color online) 2D phase domains formed by the action of spatial rocking. The
spatial period of the rocking beam is $\Lambda_{\mathrm{C}}$ = 23.2  $\mu $m 
(not seen in the images). The spatial dimensions of all subfigures are 1056 $\mu$m$\times$1266 $\mu$m.
The interferogram a) exhibits horizontal fringes, which 
are shifted in the central domain by half a period with respect to its surroundings, 
indicating a $\pi$ jump of the phase. b): amplitude map. c): phase map. d): different 
phase profiles around the vertical cut marked by a black vertical
arrow in c), evidencing $\pi$ jumps across the phase domain boundary.}
\label{domain walls}
\end{figure}

\begin{figure}[tbh]
\centerline{\includegraphics[width= 83 mm]{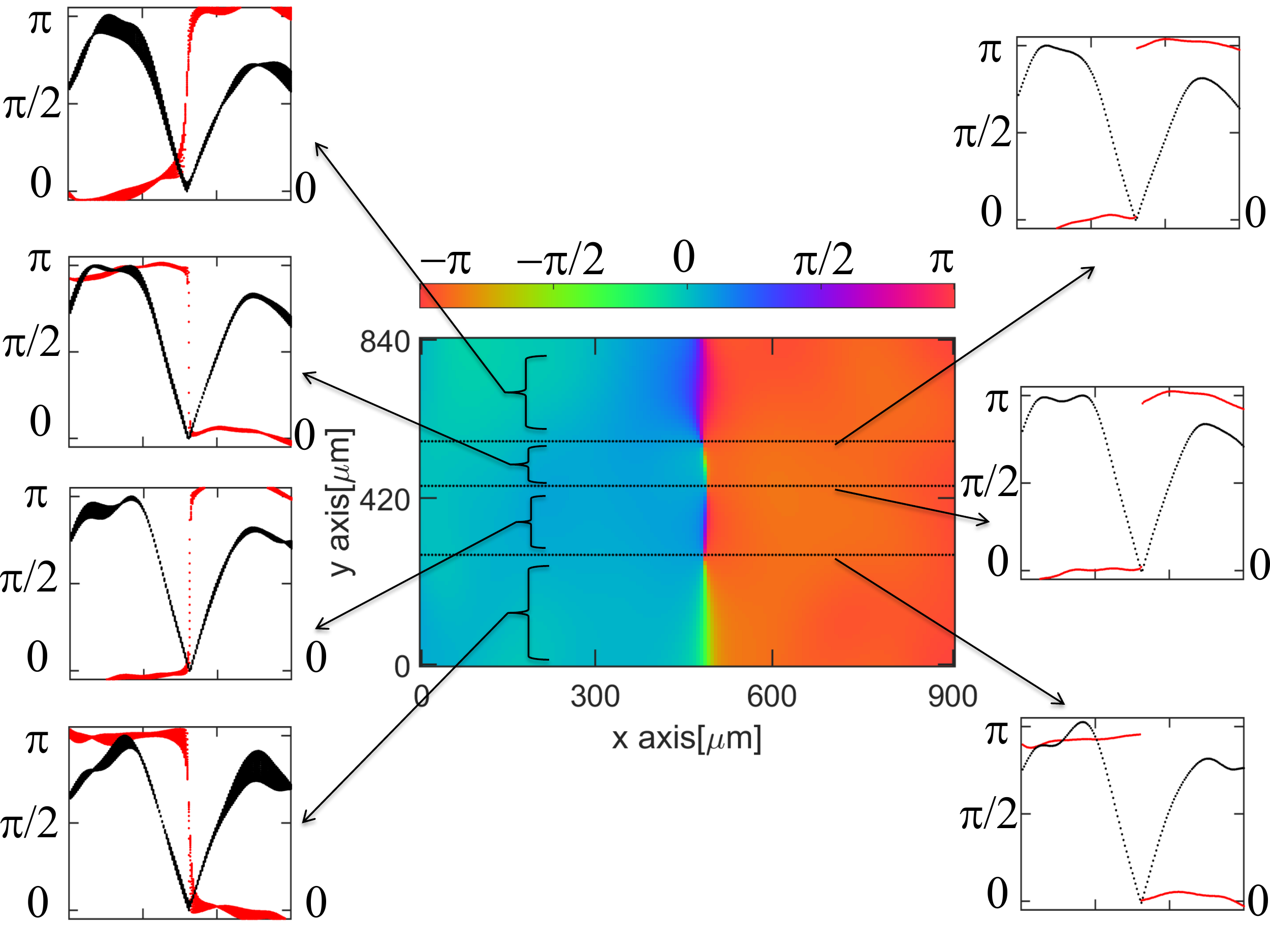}}
\caption{(Color online) Study of the chirality of a straight domain wall. Spatial
dimensions: 900 $\protect\mu $m$\times $480 $\protect\mu $m. The spatial
period of the rocking beam is $\Lambda_{\mathrm{C}} = 23.2 \protect\mu$m (not visible in the image).
The central image shows the phase map of the wall. The left column shows
different phase (red dash line) and amplitud (black dash line) profiles along to the horizontal axis, displaying $+\protect%
\pi $ or $-\protect\pi $ jumps in phase, depending on the region. The right column
shows abrupt $\protect\pi $ jumps and their respective amplitud profiles, observed at three N\'{e}el points present in the wall.}
\label{walls}
\end{figure}

Without injection our PRO is a phase invariant system (any value of the
phase is possible) which, allowed to its large Fresnel number, leads to the
spontaneous formation of optical vortices \cite{Vortex1,Vortex2}, as shown
in Fig. \ref{vortices}. These phase singularities are characterized by a
smooth rotation of the field phase by $2\pi $ on a closed loop around a core, a point
of null intensity. On the contrary, when spatial rocking is
applied, the phase invariance gets broken and now just two (opposite) phases
are preferred. In this case vortices are replaced by one-dimensional objects,
so-called phase domain walls (DWs) \cite{IB-Adolfo, IB-Coullet},
which separate spatial domains of opposite phase (Fig. \ref{domain walls}). 
The total light field is the superposition of the rocking beam and the 
beam generated by the PRO. As the latter has small spatial frequencies(small divergence angle)
as compared to the former, both contributions are well separated each other in the far
field \cite{Spatial-rocking}. In fact, the high spatial frequency of the
rocking beam (around 23.2 $\mu$m$^{-1}$) does not appear in the figures we
show here as diffraction outside the cavity filters out the large-angle components of the
rocking beam. Domain walls can be such that the phase abruptly jumps by $\pi $
across the boundary (Ising wall) or displays a relatively smooth variation (Bloch
wall), that terminology coming from solid state physics \cite{IB-Coullet}.
The phase variation across a Bloch wall can be increasing or decreasing, i.e. the phase angle 
can rotate clockwise or counter-clockwise across the wall, what means that Bloch walls are chiral \cite{IB-Adolfo,IB-Coullet} 
(see Fig. \ref{domain walls}d.). Althought we demonstrate the existence of DWs induced by spatial rocking (see Fig. \ref{domain walls}.), the curvature of DWs can induce an additional dynamics, which should be removed in order to separate the effects. Thus special efforts are taken to ensure a quasi-one dimensional regime in the transverse plane \cite{IB-Adolfo} in order to avoid that influence of DW curvature in the observed dynamics \cite{Gomila1, Gomila2}. Quasi-one dimensionality is achieved by placing slits inside the nonlinear cavity, at the Fourier planes of telescopes T2 and T3 (see Fig. \ref{scheme}.). The width of the slits is adjusted to the size of the diffraction spot in these planes. In this way waves with too large inclination are not compatible with the
diffraction constraints of the cavity. We also use another pair of slits in the near field, i.e. close to the cavity mirrors (MIRROR
in Fig. \ref{scheme}.), which allows to select specific parts of the nonlinear crystal, , e.g. particularly homogeneous regions of the crystal 
(see Fig. \ref{walls}.). In quasi one-dimensional systems the interface separating phase
domains can display positive chirality in a particular segment, and negative chirality
in the adjacent one, as shown in Fig. \ref{walls}. By continuity,
between the sections of Bloch walls with opposite chirality a point of null
chirality (a kind of Ising wall) appears. This point is a kind of bound
vortex, which is the nonequilibrium analogue of a N\'{e}el point in solid
state physics \cite{Neel}.
Usually nonequilibrium Bloch walls move according to their chirality \cite{IB-Adolfo, IB-Coullet}, so that opposite chirality walls move in opposite directions. When two Bloch walls with opposite chirality are separated by a N\'{e}el point, such as in Fig. \ref{walls}, such motion generically leads to the emergence of spiral waves with center at the (static) N\'{e}el point \cite{Spiral1,Spiral2}. In our case such effect has not been observed, due to two different causes: (i) the dynamics of the system is very slow (in previous experiments on the Ising-Bloch transition in PROs the velocity of the Bloch walls was measured to be on the order of 1 $\mu m$/second) and we did not perform long-time observations, and (ii) the quasi 1D geometry affects the possible wall motion because walls ending at a boundary are not free to move but are always perpendicular to it \cite{Steering1}.
\\
Concluding in this work we have given a first experimental evidence of the phase bistability
appearing in a large Fresnel number laser-like system submitted to spatial rocking 
\cite{Spatial-rocking}. The large Fresnel number of the cavity allows the formation of
phase patterns, which take the form of domain walls due to phase bistability imposed by
spatial rocking. Such phase bistable spatial structures can be efficiently written, erased, and moved
across the transverse section of the system , as has been demonstrated in temporal
rocking \cite{Steering1, Steering2}. The reported theoretical results 
thus indicate that the recent predictions on the excitation of cavity solitons by spatial rocking in broad 
area semiconductor, lasers \cite{semiconductor} and vertical surface emission lasers
\cite{Rock-VCSEL} open the way to new types of optical information processing in such 
semiconductor micro lasers \cite{SCS1,SCS2,SCS3}, based on phase-bistable cavity solitons.\\
We are indebted to Javier Garc\'{\i}a-Monreal and Mart\'{\i}n Sanz-Sabater
(Departament d'\`{O}ptica, Universitat de Val\`{e}ncia) for their help and
advice. This work was supported by the Spanish Ministerio de Econom\'{\i}a y
Competitividad and the European Union FEDER (project No. FIS2011-26960) and 
FIS2011-29734-C02-01.

\end{document}